\documentclass[12pt]{article}
\usepackage{amssymb,amsmath,epsfig}

\numberwithin{equation}{section}

\begin{document}

\title{\textbf{Fermions Tunneling from Charged anti-de Sitter Black Holes}}

\author{M. Sharif \thanks{msharif.math@pu.edu.pk} and Wajiha Javed
\thanks{wajihajaved84@yahoo.com} \\
Department of Mathematics, University of the Punjab,\\
Quaid-e-Azam Campus, Lahore-54590, Pakistan.}

\date{}
\maketitle

\begin{abstract}
We study Hawking radiation as a phenomenon of tunneling through
event horizons of charged torus-like as well as dilaton black holes
involving cosmological constant based on Kerner and Mann's
formulation. We obtain tunneling probabilities as well as Hawking's
emission temperature of outgoing charged particles by applying the
semiclassical WKB approximation to the general covariant Dirac
equation. The graphical behavior of Hawking temperature and horizon
radius is investigated. We find results consistent with those
already given in literature.
\end{abstract}
{\bf Keywords:} Quantum tunneling; Torus-like and dilaton black holes.\\
{\bf PACS numbers:} 04.70.Dy; 04.70.Bw; 11.25.-w

\section{Introduction}

Hawking \cite{A1A} demonstrated that the surface area of event
horizon of a black hole (BH) can never decrease with time. This led
to an interesting new era of research in BH physics. Bekenstein
\cite{A2A} proved that the BH entropy is proportional to its horizon
area. This made a foundation of the correspondence between the
classical as well as BH laws of thermodynamics. Thus, BH refers to
as a thermodynamic substance having entropy corresponding to its
horizon area, a temperature related to its surface gravity $\kappa$
and an internal energy comparable to its mass. According to Hawking
\cite{A4A}, a BH is not sufficiently black but actually discharges
radiation named as Hawking radiation.

Semiclassically, quantum field distribution in the environment of a
BH manifests that it emits black body radiation with thermal
spectrum at Hawking temperature $T_H=\frac{\kappa}{2\pi}$ and
entropy $S=A/4$. Quantum gravity encourages the phenomenon of
information loss from BHs, according to which the Hawking
temperature increases as the mass of the BH diminishes. Thus, a BH
would radiate repeatedly some of its mass through Hawking radiation,
leading to increase in Hawking temperature. Since BH radiation is a
continuous process, so with the passage of time, finally it
evaporates completely. The final stage of the evaporated BH has
considerably small size of the order of Planck length. Consequently,
the ultimate stage of evaporation can only be illustrated by a
quantum gravity theory.

The radiation spectrum of BHs has been discussed by using various
techniques. Damour and Ruffini \cite{S4} provided unification of
quantum mechanics and general relativity to analyze Hawking's
radiation progression. Kraus and Wilczek \cite{S6} extended the
tunneling procedure to investigate the Hawking radiation by
discussing correlations between incoming particles and the
radiation. They considered self-gravitational effects of radiation
and found that particles no longer move along geodesics and
corresponding radiation spectrum is not precisely thermal.
Keski-Vakkuri and Kraus \cite{S7} developed a tunneling algorithm to
study the radiation spectrum based on the WKB approximation together
with complex time path techniques. This algorithm is used to
calculate the tunneling probabilities for systems with time
dependent background. Later, this method was used by Parikh and
Wilczek \cite{S8}-\cite{S10} in order to formulate another tunneling
approach to investigate the tunneling radiation features of the
Schwarzschild and Reissner$-$Nordstr\"{o}m (RN) BHs.

Following Parikh and Wilczek semiclassical tunneling method, there
has been a debate for tunneling of particles across the horizon
either massive or massless as well as charged or uncharged. Hemming
and Keski-Vakkuri \cite{S11} explored the radiation spectrum from
anti-de Sitter BH, while Medved \cite{S12} studied from a de Sitter
BH. Zhang and Zhao \cite{S13,S14} investigated radiation phenomenon
from an axisymmetric BH. These results hold the Parikh and Wilczek
conclusion, i.e., the true Hawking radiation spectrum is not purely
thermal due to the significance of self-gravitational effects.
Kerner and Mann \cite{S15} proposed a method to examine fermion's
tunneling by extending the idea that the Hawking radiation may emit
fermions. For this purpose, they used the general Dirac equation to
investigate the radiated spin particle's action.

Recently, we have explored some work \cite{W} about the tunneling
phenomenon for different BHs by using the above referred techniques.
Here, we investigate the radiation spectrum via fermions tunneling
of charged particles through the horizon of a BH having torus-like
topology as well as charged dilatonic BHs and also the corresponding
Hawking temperature. The graphical representation of Hawking
temperature and horizon radius is also given. We discuss the effects
of cosmological constant and charge on temperature as well as
horizon radius. The format of the paper is the following: Section
\textbf{2} provides review of the basic procedure about tunneling
spectrum. In section \textbf{3}, we investigate fermions tunneling
spectrum to both torus-like BH and dilaton BH in anti-de Sitter
spaces. Finally, section \textbf{4} summarizes the results.

\section{Charged Fermions Tunneling}

In order to discuss tunneling spectrum of the given BHs, we adopt a
semiclassical tunneling method based on the Kerner and Mann
technique \cite{S15}. The Dirac equation for charged fermions in
covariant coordinate system is \cite{rn}
\begin{equation}
\imath\gamma^\mu\left(D_\mu-\frac{\imath
q}{\hbar}A_\mu\right)\Psi+\frac{m}{\hbar}\Psi=0,\quad
\mu=0,1,2,3\label{2'}
\end{equation}
where $m$ and $q$ denote mass and charge of the emitted fermion
particle, respectively, $A_\mu$ is the 4-potential, $\Psi$ is the
wave function and
\begin{equation*}
D_\mu=\partial_\mu+\Omega_\mu,\quad
\Omega_\mu=\frac{1}{2}\imath\Gamma^{\sigma\delta}_
\mu\Sigma_{\sigma\delta},\quad\Sigma_{\sigma\delta}=
\frac{1}{4}\imath[\gamma^\sigma,\gamma^\delta].
\end{equation*}
The antisymmetric property of the Dirac matrices \cite{ks1}, i.e.,
$[\gamma^\sigma,\gamma^\delta]=0$ for $\sigma=\delta$ and
$[\gamma^\sigma,\gamma^\delta]=-[\gamma^\delta,\gamma^\sigma]$ for
$\sigma\neq\delta$, reduces the Dirac equation (\ref{2'}) in the
following form
\begin{equation}
\imath\gamma^\mu\left(\partial_\mu-\frac{\imath
q}{\hbar}A_\mu\right)\Psi+\frac{m}{\hbar}\Psi=0.\label{2}
\end{equation}

The spinor wave function $\Psi$ (related to the particle's action)
has two spin states: spin-up and spin-down in $+$ve and $-$ve
$r$-directions, respectively. For the spin-up and spin-down particle
solution, we assume \cite{S15}
\begin{eqnarray}
\Psi_\uparrow(t,r,\varphi,\psi)=\left[\begin{array}{c}
E(t,r,\varphi,\psi)\\0\\
F(t,r,\varphi,\psi)\\0
\end{array}\right]\exp\left[\frac{\imath}{\hbar}
I_\uparrow(t,r,\varphi,\psi) \right],\\\label{1}
\Psi_\downarrow(t,r,\varphi,\psi)=\left[\begin{array}{c}
0\\G(t,r,\varphi,\psi)\\0\\
H(t,r,\varphi,\psi)
\end{array}\right]\exp\left[\frac{\imath}{\hbar}
I_\downarrow(t,r,\varphi,\psi) \right],\label{}
\end{eqnarray}
where $I_{\uparrow/\downarrow}$ symbolize the radiated particle's
action regarding spin-up/spin-down, respectively. The particle's
action is described by
\begin{equation}
I_\uparrow=-\omega t+W(r)+J(\varphi,\psi),\label{WWWW}
\end{equation}
where $\omega,~J$ and $W$ are energy, angular momentum and arbitrary
function, respectively.

The line element for torus-like BH can be written in the following
form
\begin{equation}
\textmd{d}s^2=-g(r)\textmd{d}t^2
+\frac{\textmd{d}r^2}{g(r)}+r^2(\textmd{d}\varphi^2+\textmd{d}\psi^2).\label{M1}
\end{equation}
Using the ansatz (\ref{WWWW}) into the Dirac equation with
$E=-\imath F,~E=\imath F$ and Taylor's expansion of $g(r)$ near the
event horizon $r_+$, it follows that
\begin{eqnarray}
-F\left[\frac{-(\omega+qA_t)}{\sqrt{(r-r_+)g^\prime(r_+)}}
+\sqrt{(r-r_+)g^\prime(r_+)}W^\prime(r)\right]+mE=0,\label{11}\\
-\frac{F}{r}\left[\partial_\varphi J+\imath\partial_\psi J\right]=0,\label{12}\\
E\left[\frac{-(\omega+qA_t)}{\sqrt{(r-r_+)g^\prime(r_+)}}
-\sqrt{(r-r_+)g^\prime(r_+)}W^\prime(r)\right]+mF=0,\label{13}\\
-\frac{E}{r}\left[\partial_\varphi J+\imath\partial_\psi
J\right]=0.\label{14}
\end{eqnarray}
It is mentioned here that, the condition $E=-\imath F$ holds for
outgoing solutions, while $E=\imath F$ for the incoming solutions.
Equations (\ref{12}) and (\ref{14}) imply that
\begin{equation}
J(\varphi,\psi)=c_1e^{c_2(\varphi+\imath\psi)},
\end{equation}
where $c_1$ and $c_2$ are arbitrary constants. This expression can
be used in Eq.(\ref{WWWW}) to calculate particle's action. For the
massless case ($m=0$), Eqs.(\ref{11}) and (\ref{13}) yield
\begin{equation}
W^\prime(r)=W^\prime_+(r)=-W^\prime_-(r)=\frac{\omega+qA_t}{(r-r_+)
g^\prime(r_+)}.\label{AW}
\end{equation}
Integrating this equation with respect to $r$, we get
\begin{equation}
W_+(r)=-W_-(r)=\frac{\omega+qA_t}{g^\prime(r_+)}\int\frac{dr}{(r-r_+)},\label{AW1}
\end{equation}
where $W_+$ and $W_-$ correspond to the outgoing and incoming
solutions, respectively. This equation represents the pole at the
horizon $r=r_+$.

\section{Charged anti-de Sitter Black Holes}

In this section, we examine fermions tunneling spectrum as radiation
process from charged anti-de Sitter BHs essentially torus-like and
dilaton BHs.

\subsection{Torus-like Black Hole}

In general relativity, charged rotating BHs are described by mass
$M$, angular momentum $J$ and electric charge $Q$. The topology of
spacelike solution of the event horizon of these BH solutions is a
two-dimensional sphere $S^2$. It would be interesting to study such
BH solutions of the Einstein-Maxwell field equations whose event
horizons have topologies other than $S^2$, e.g., torus-like. We
explore charged particles tunneling from the torus-like BH solution
with negative cosmological constant.

The torus-like BH solution is characterized by the line element
\cite{torus1995}
\begin{eqnarray}\label{W1}
\textmd{d}s^2&=&-\left(-\frac{1}{3}\Lambda r^2-\frac{2M}{\pi
r}+\frac{4Q^2}{\pi
r^2}\right)\textmd{d}t^2+\left(-\frac{1}{3}\Lambda r^2-\frac{2M}{\pi
r}+\frac{4Q^2}{\pi r^2}\right)^{-1}dr^2\nonumber\\
&+&r^2(\textmd{d}\varphi^2+\textmd{d}\psi^2),
\end{eqnarray}
where $\varphi,\psi\in[0,2\pi]$. Equations (\ref{M1}) and (\ref{W1})
lead to
\begin{equation}
g(r)=-\frac{1}{3}\Lambda r^2-\frac{2M}{\pi r}+\frac{4Q^2}{\pi r^2}.
\end{equation}
This line element will be singular for $g(r)=0$, i.e.,
\begin{equation}
-\frac{1}{3}\Lambda r^2-\frac{2M}{\pi r}+\frac{4Q^2}{\pi
r^2}=0.\label{W2}
\end{equation}
When $\Lambda>0$, the only non-zero horizon radius is
\begin{equation}
r_h=\frac{1}{2}[\Upsilon^{1/2}+\{-\Upsilon+2[\Upsilon^2+
(48/\pi\Lambda)Q^2]^{1/2}\}^{1/2}],
\end{equation}
for any $M$ and $Q$, where $\Upsilon$ is given by
\begin{eqnarray}
\Upsilon=\left\{\frac{(6M)^2}{2\pi^2\Lambda^2}+[(\frac{(6M)^2}{2\pi^2
\Lambda^2})^2+(\frac{16}{\pi\Lambda}Q^2)^2]^{1/2}\right\}^{1/3}\nonumber\\
+\left\{\frac{(6M)^2}{2\pi^2\Lambda^2}-[(\frac{(6M)^2}{2\pi^2\Lambda^2})^2
+(\frac{16}{\pi\Lambda}Q^2)^2]^{1/2}\right\}^{1/3}.\label{WW}
\end{eqnarray}
Incorporating the fact that the line element (\ref{W1}) is not
static for large $r$, it implies that the metric with positive
$\Lambda$ does not describe a BH solution.

For $\Lambda<0$, Eq.(\ref{W2}) has two positive solutions
\begin{equation}
r_\pm=\frac{1}{2}\left[\Upsilon^{1/2}\pm\{-\Upsilon+2[\Upsilon^2+
(48/\pi\Lambda)Q^2]^{1/2}\}^{1/2}\right]\label{1WW1}
\end{equation}
as long as
\begin{equation*}
0\leq Q^2\leq\frac{3}{8}(3M^4/2\pi|\Lambda|)^{1/3}.
\end{equation*}
Thus, the metric has coordinate singularities at the horizon radii
$r_\pm$ for which
\begin{equation}
-\frac{1}{3}\Lambda r_\pm^2-\frac{2M}{\pi r_\pm}+\frac{4Q^2}{\pi
r_\pm^2}=0,
\end{equation}
where $r_+$ and $r_-$ represent outer and inner horizons of the BH.
The only nonvanishing ingredient of electromagnetism in the Dirac
equation is $A_t=-\frac{4Q}{r}$.

For semiclassical tunneling probability, the wave equation is
multiplied by its complex conjugate. In this way, the part of curved
route (trajectory) of particles (that originate from outside region
of the BH and prolonged to the observer) will not supplement to the
calculation of the absolute tunneling probability and can be
neglected. Thus, the only sector of the wave (trajectory) that just
provides the tunneling probability is the contour encompassing the
BH horizon. We choose that the outgoing particle contour is in the
lower half plane, so by using mathematically equivalent convention
we do not multiply the contour integration by a negative sign
\cite{rbmann}.

Inserting the value of $g$ in Eq.(\ref{AW1}), we obtain
\begin{eqnarray}
W_+(r)=-W_-(r)=\frac{\omega+qA_t}{-\frac{2}{3}\Lambda
r_++\frac{2M}{\pi r_+^2}-\frac{8Q^2}{\pi
r_+^3}}\int\frac{dr}{r-r_+}.
\end{eqnarray}
Integrating around the pole, it follows that
\begin{equation}
W_+(r)=-W_-(r)=\frac{\pi \imath(\omega+qA_t)}{-\frac{2}{3}\Lambda
r_++\frac{2M}{\pi r_+^2}-\frac{8Q^2}{\pi r_+^3}},
\end{equation}
where the imaginary parts of $W_+$ and $W_-$ yield
\begin{equation}
\textmd{Im}W_+=-\textmd{Im}W_-=\frac{\pi(\omega+qA_t)}{-\frac{2}{3}\Lambda
r_++\frac{2M}{\pi r_+^2}-\frac{8Q^2}{\pi r_+^3}}.
\end{equation}
Thus, the particle's tunneling probability from inside of the event
horizon to outside is
\begin{eqnarray}
\Gamma=\frac{\textmd{Prob}[\textmd{out}]}
{\textmd{Prob}[\textmd{in}]}= \exp[-4\textmd{Im}W_+]
=\exp\left[-\frac{2\pi(\omega+qA_t)} {-\frac{1}{3}\Lambda
r_++\frac{M}{\pi r_+^2}-\frac{4Q^2}{\pi r_+^3}}\right].\label{123}
\end{eqnarray}
\begin{figure}\center
\epsfig{file=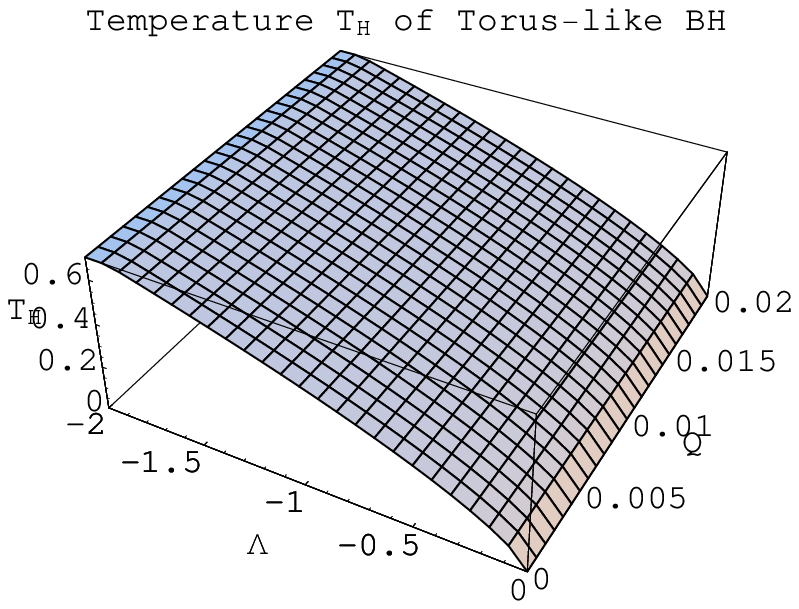, width=0.50\linewidth}\\
\caption{Hawking temperature $T_H$ versus cosmological constant
$\Lambda$ and charge $Q$}
\epsfig{file=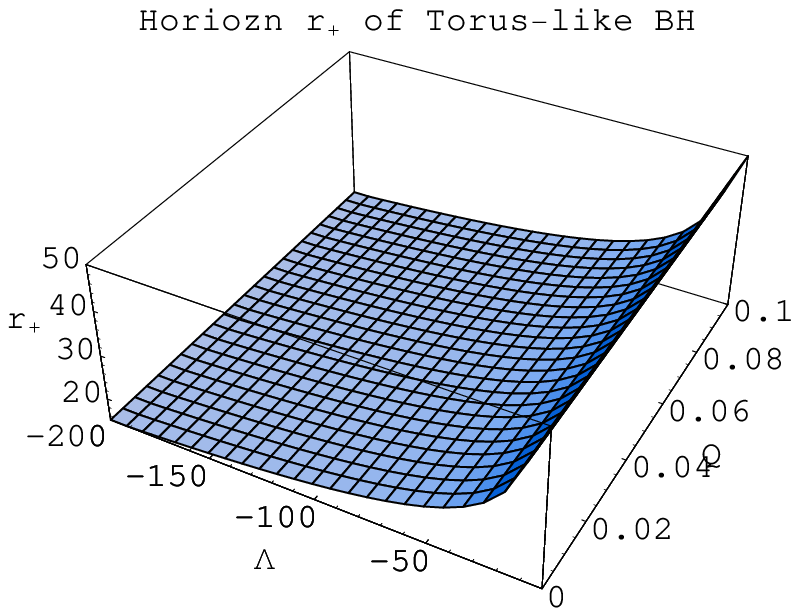, width=0.50\linewidth}\\
\caption{Horizon radius $r_+$ versus cosmological constant $\Lambda$
and charge $Q$}
\end{figure}

We expand the action in terms of particle energy $\omega$ so that
the linear order leads to the Hawking temperature. The higher order
terms represent the self-interaction effects resulting from the
energy conservation. Thus, the emission rate in the tunneling
approach just to first order in $\omega$ retrieves the Boltzmann
factor, $\exp[-\beta \omega]$, where $\beta=\frac{1}{T_H}$
\cite{s2}. It follows that the emission rate in the high energy is
proportional to the Boltzmann factor, i.e.,
\begin{equation}
\Gamma\simeq\exp\left[-\frac{2\pi \omega} {-\frac{1}{3}\Lambda
r_++\frac{M}{\pi r_+^2}-\frac{4Q^2}{\pi r_+^3}}\right].\label{122}
\end{equation}
Comparing with $\Gamma\simeq\exp[-\beta \omega]$, the Hawking
temperature takes the form
\begin{equation}
T_H=\frac{1}{2\pi}\left(-\frac{1}{3}\Lambda r_++\frac{M}{\pi
r_+^2}-\frac{4Q^2}{\pi r_+^3}\right)\label{NN5}.
\end{equation}
For the massive case ($m\neq0$), following the same steps, we can
obtain the same temperature. Thus, the behavior of massive particles
near the BH horizon is the same as that for the massless particles.

The behavior of Hawking temperature (\ref{NN5}) of torus-like BH in
anti-de Sitter spacetime for $M=100$ (based on the cosmological
constant $\Lambda$ and electric charge $Q$) is shown in Figure
\textbf{1}. This indicates that torus-like BH temperature increases
as $\Lambda$ decreases, while charge has no effect. In this
background, the location of the horizon radius can be obtained by
using Eqs.(\ref{WW}) and (\ref{1WW1}). We also plot the horizon
radius with cosmological constant $\Lambda$ and charge $Q$ for
$M=100000$ displayed in Figure \textbf{2}. This shows that the
horizon radius decreases gradually as $\Lambda$ decreases. These
results are consistent with those given in \cite{D1}.

\subsection{Dilaton Black Holes}

Dilaton is characterized by a scalar field occurring in the string
theory with low energy limit. Dilaton field has considerable outcome
on causal structure and thermodynamics of the BH. The standard form
of the line element for the static spacetime can be prescribed as
\cite{DILA}
\begin{equation}
ds^2=-g(r)dt^2+\frac{1}{g(r)}dr^2+f(r)^2d\Omega^2_{k,2},\label{M4}
\end{equation}
where $d\Omega^2_{k,2}$ is the line element for a two-dimensional
hypersurface with constant curvature
\begin{eqnarray}
d\Omega^2_{k,2}=\left\{\begin{array}{cc}
d\theta^2+\sin^2\theta d\varphi^2 &\textmd{for}~k=1, \\
d\theta^2+\theta^2d\varphi^2 &\textmd{for}~k=0, \\
d\theta^2+\sinh^2\theta d\varphi^2&\textmd{for}~k=-1.
\end{array}\right.\label{9}
\end{eqnarray}
The Maxwell equation,
$\partial_\mu(\sqrt{-g}e^{2a\phi}F^{\mu\nu})=0$, gives the only
non-zero component of the Maxwell field tensor as
$F_{01}=\frac{Qe^{2a\phi}}{f^2}$. The dilaton field $\phi$ can be
defined as \cite{PLB}
$e^{2a\phi}=(1-\frac{r_-}{r})^{\frac{2a^2}{(1+a^2)}}e^{-2a\phi_0}$
with $e^{2a\phi_0}=\frac{r_+r_-}{(1+a^2)Q^2}$ \cite{D1}, where
$\phi_0$ is the dilaton field at infinity and $a$ is an arbitrary
parameter which represents the intensity of the coupling of the
Maxwell field to the dilaton field.

The topological BH solution can be obtained as \cite{DILA}
\begin{eqnarray}
g&=&\left(k-\frac{r_+}{r}\right)\left(1-\frac{r_-}{r}\right)^
{\frac{1-a^2}{1+a^2}}-\frac{1}{3}\Lambda
r^2\left(1-\frac{r_-}{r}\right)^{\frac{2a^2}{1+a^2}},\nonumber\\
Q^2&=&\frac{r_+r_-}{1+a^2},\quad 2M=r_++\frac{1-a^2}{1+a^2}r_-,\quad
f=r\left(1-\frac{r_-}{r}\right)^{\frac{a^2}{1+a^2}}.\label{M6}
\end{eqnarray}
Equation (\ref{M6}) implies that either $a=0$ or $\phi_0=constant$.
For $\phi_0=constant$, the corresponding dilaton field is
\begin{equation}
e^{2a\phi}=(1-\frac{r_-}{r})^{\frac{2a^2}{1+a^2}}.
\end{equation}
The locations of the event horizon $r_+$ and inner horizon $r_-$ are
\begin{equation}
r_\pm=\frac{1+a^2}{1\pm a^2}[M\pm\sqrt{M^2-(1-a^2)Q^2}].
\end{equation}
The Dirac equation with the spinor wave function leads to the same
equations as given by Eqs.(\ref{11}) and (\ref{13}) with $g(r)$
defined in Eq.(\ref{M6}). The corresponding value of
$J(\theta,\varphi)$ to obtain particle's action is found to be
\begin{equation}
J(\theta,\varphi)=c_3e^{c_4[-f(r)\ln\mid\csc\theta-\cot\theta\mid+\imath\varphi]},
\end{equation}
where $c_3$ and $c_4$ are arbitrary constants.

The tunneling probability is given by
\begin{equation*}
\Gamma=\exp[-4\textmd{Im}W_+]
\end{equation*}
which implies that
\begin{equation}
\Gamma=\exp[-\frac{4\pi(\omega+qA_t)}{g^\prime(r_+)}]\simeq\exp[-\beta
\omega].
\end{equation}
Consequently, we have
\begin{equation}
T_H=\frac{g^\prime(r_+)}{4\pi}.\label{A13}
\end{equation}
The horizon radii can be obtained by using $g(r)=0$, which means
that
\begin{equation}
(k-\frac{r_+}{r})(1-\frac{r_-}{r})^{-\frac{2a^2}{1+a^2}}=
\frac{\Lambda}{3}r^2(1-\frac{r_-}{r})^{-\frac{(1-a^2)}{1+a^2}}.\label{2WW2}
\end{equation}
Thus, we can write
\begin{eqnarray}
g^\prime(r)&=&\frac{\Lambda}{3(1+a^2)}r_-
r^{\frac{1-a^2}{1+a^2}}(r-r_-)^{\frac{a^2-1}{a^2+1}}(1-3a^2)
\nonumber\\&+&
r_+r^{\frac{-(3+a^2)}{1+a^2}}(r-r_-)^{\frac{1-a^2}{1+a^2}}-
\frac{2\Lambda}{3}r^{\frac{1-a^2}{1+a^2}}
(r-r_-)^{\frac{2a^2}{1+a^2}}.
\end{eqnarray}
Inserting this value of $g^\prime(r)$ in Eq.(\ref{A13}), the Hawking
temperature can be found by using spin-up fermions tunneling from
the dilaton BH as
\begin{eqnarray}
T_H&=&\frac{1}{4\pi}\left[\frac{\Lambda}{3}(M-\sqrt{M^2-(1-a^2)Q^2})
\left\{1-\frac{1}{r_+}\left(\frac{1+a^2}{1-a^2}\right)(M\nonumber\right.
\right.\\&-&\left.\left.\sqrt{M^2-(1-a^2)Q^2})\right\}
^{\frac{a^2-1}{a^2+1}}+\frac{1}{r_+^2}(M+\sqrt{M^2-(1-a^2)Q^2})
\nonumber\right.\\&\times&\left.\left\{1-\frac{1}{r_+}\left(\frac{1+a^2}{1-a^2}\right)
(M-\sqrt{M^2-(1-a^2)Q^2})\right\}^{\frac{1-a^2}{1+a^2}}\nonumber\right.\\
&-&\left.\frac{2\Lambda}{3}\left(\frac{a^2}{1-a^2}\right)(M-\sqrt{M^2-(1-a^2)Q^2})
\nonumber\right.\\&\times&\left.\left\{1-\frac{1}{r_+}\left(\frac{1+a^2}{1-a^2}\right)
(M-\sqrt{M^2-(1-a^2)Q^2})\right\}^{\frac{a^2-1}{a^2+1}}\nonumber\right.\\&-&\left.
\frac{2\Lambda}{3}r_+\left\{1-\frac{1}{r_+}\left(\frac{1+a^2}{1-a^2}\right)
(M-\sqrt{M^2-(1-a^2)Q^2})\right\}^{\frac{2a^2}{1+a^2}}\right].\nonumber\\\label{M3}
\end{eqnarray}
For spin-down particles, the same Hawking temperature can be
recovered as given above.
\begin{figure}\center
\epsfig{file=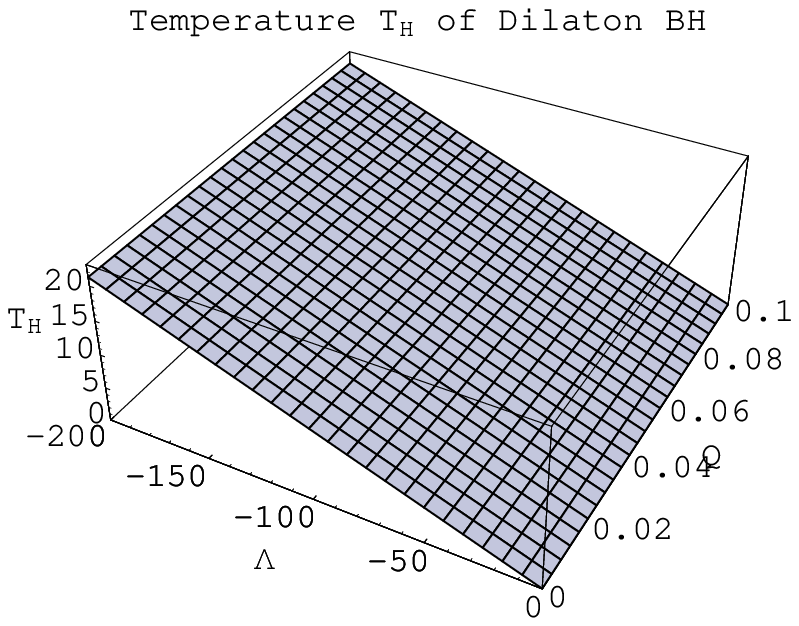, width=0.50\linewidth}\\
\caption{For $0\leq a<1$, the Hawking temperature $T_H$ versus
cosmological constant $\Lambda$ and charge $Q$}
\end{figure}
\begin{figure}\center
\epsfig{file=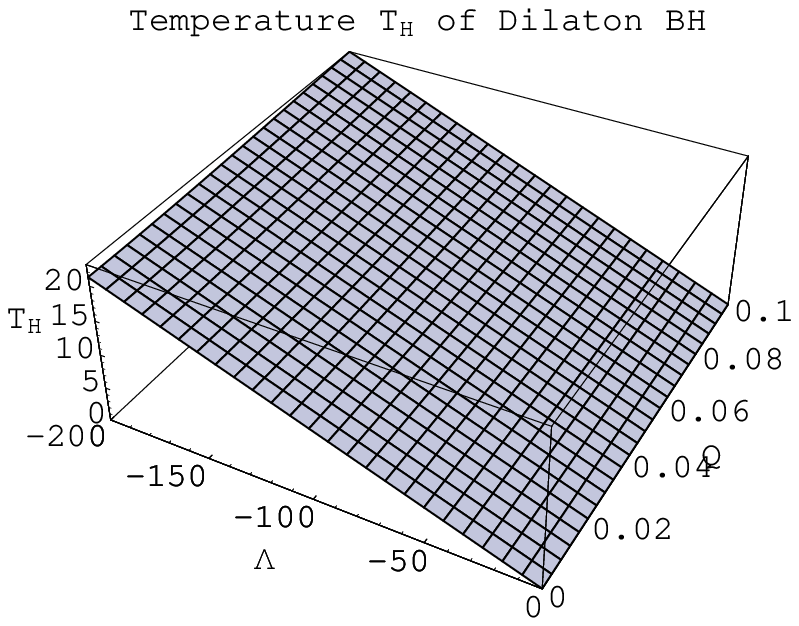, width=0.50\linewidth}\\
\caption{For $a>1$, the Hawking temperature $T_H$ versus
cosmological constant $\Lambda$ and charge $Q$}
\epsfig{file=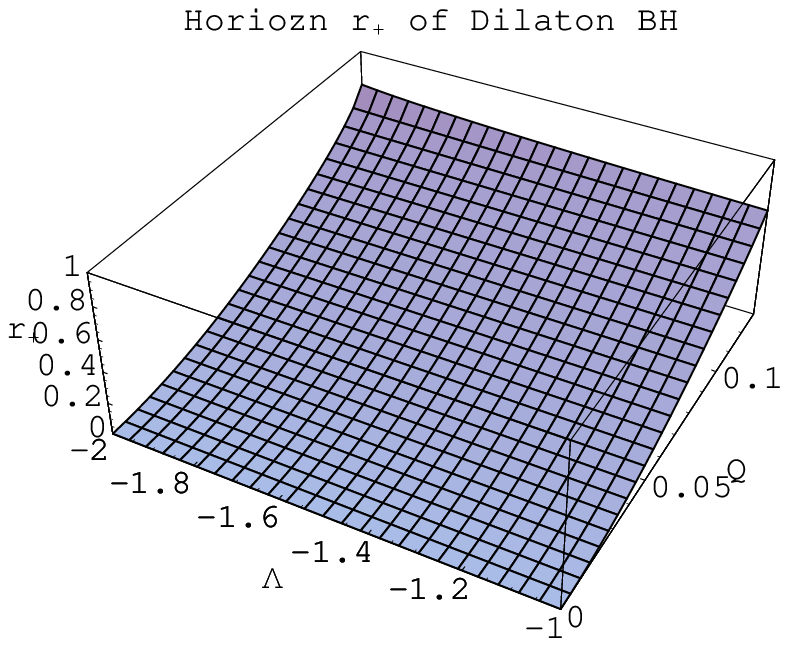, width=0.50\linewidth}\\
\caption{For $a>1$, the horizon radius $r_+$ versus cosmological
constant $\Lambda$ and charge $Q$}
\end{figure}

The Hawking temperature (\ref{M3}) of fermions tunneling through the
event horizon of dilaton anti-de Sitter BHs is found to be
consistent with that given in \cite{D1}. For $\Lambda=0$, it reduces
to the Hawking temperature corresponding to the charged dilatonic
BHs \cite{D2}. It is worth mentioning here that the Hawking
temperature for dilaton anti-de Sitter BHs with arbitrary coupling
constant $a$ is independent of $k$ by using Eq.(\ref{2WW2}). Thus,
the Hawking temperature of fermions tunneling through surfaces of
event horizons of two-dimensional sphere (for $k=1$),
two-dimensional torus (for $k=0$) and two-dimensional hyperboloid
turns out to be the same.

For $a=0=\Lambda$ and $k=1$, the line element (\ref{M4}) reduces to
the RN BH and also the Hawking temperature $T_H$ (\ref{M3}) reduces
to the RN BH temperature \cite{S8}
\begin{equation}
T_H=\frac{1}{2\pi}\frac{\sqrt{M^2-Q^2}}{(M+\sqrt{M^2-Q^2})^2}.
\end{equation}
The condition for the extremal BH, i.e., $r_+=r_-$, suggests that
the surface gravity and the Hawking temperature of dilaton BH vanish
for all values of the coupling constant $a$. We have plotted the
effect of $a$ on Hawking temperature. For $0\leq a<1$ and $M=1$, the
Hawking temperature (\ref{M3}) increases as $\Lambda$ decreases and
hence it diverges as shown in Figure \textbf{3}. For $a>1,~T_H$
diverges which leads to the extremal BH case \cite{D2}. In this
case, we assume $M=1$ and $a=10$ to explore $T_H$ which is related
to the cosmological constant $\Lambda$ and electric charge $Q$. The
graphical expression of $T_H$ is given in Figure \textbf{4}. Thus,
the Hawking temperature of dilaton BH increases as $\Lambda$
decreases but charge increases. This leads to the fact that in
anti-de Sitter spacetime, the negative value of $\Lambda$ represents
the force of attraction supporting gravitational collapse. Thus, the
temperature increases and finally diverges.

The horizon equation $g(r)=0$ ($g(r)$ is given in (\ref{M6})) is
found to be
\begin{eqnarray}
&-&\frac{\Lambda}{3}r^4+kr^2+r[-(M-\sqrt{M^2-(1-a^2)Q^2})\nonumber\\
&-&k\frac{1+a^2}{1-a^2}(M-\sqrt{M^2-(1-a^2)Q^2})]+(1+a^2)Q^2=0\label{M5}
\end{eqnarray}
by assuming that the dilaton field takes its extreme value,
$(1-\frac{r_-}{r})^{\frac{2a}{(1+a^2)}}=\frac{r_+r_-}{(1+a^2)Q^2}$.
When the coupling parameter $a=0$ (or $\phi=constant=0$), the
potential of the dilaton gravity theory is reduced to the
cosmological constant. Also, we formulate the graphical behavior of
the horizon radius (\ref{M5}) with cosmological constant $\Lambda$
and charge $Q$ in the dilaton field extremal limit, $\phi=\phi_0$,
shown in Figure \textbf{5}. We have taken $M=1,~a=10$ and for all
$k=-1,0,1$, it turns out that the horizon radius increases in a
gradual manner as $\Lambda$ decreases. Thus for $a>1,~\Lambda$
decreases while the horizon radius shows opposite behavior with that
given in \cite{D1} and leads to the RN-de Sitter case \cite{new}.
The BH still radiates but does not vanish, i.e., as horizon radius
increases, the charge also increases.

This shows rapid expansion in the horizon radius of the dilaton BHs
which proves the fact that when $\phi=\phi_0=0$, the charged dilaton
BHs lead to the RN-de Sitter BH solution \cite{new}. Thus, the
dilaton field at infinity does not support the phenomenon of BH
evaporation \cite{D2}. For $a=1$, Eqs.(\ref{M3}) and (\ref{M5})
imply that there exists no graphical representation of the Hawking
temperature as well as horizon radius (extremal case). For $0\leq
a<1$, the graphical illustration of horizon radius provides that
$\Lambda$ and $Q$ constantly affect the horizon radius and it
vanishes for all $\Lambda\leq-0.1$ and $0\leq Q\leq0.2$.

\section{Discussions}

Semiclassically, from a pair of particles (created in the vicinity
of the event horizon due to vacuum fluctuations), a positive energy
particle has the ability to tunnel outside the horizon (contradicts
classical approach), while a negative energy particle tunnels
inward. Thus quantum mechanically, horizon plays a role of two way
energy barrier for a pair of positive and negative energy particles.
We have considered tunneling probabilities for both incoming as well
as outgoing particles. Relating these tunneling probabilities with
the expression, $\exp[-\beta \omega]$, we can recover the
corresponding Hawking temperature for these BHs at event horizons.

To this end, we have used Kerner and Mann semiclassical fermions
tunneling process through WKB approximation. We have explored the
process of charged fermions tunneling from charged torus-like and
dilaton anti-de Sitter BHs. We have recovered the corresponding
Hawking temperatures of these BHs through this tunneling process.
The graphical behavior of horizon radius and temperature is also
given. For the dilaton BH solution, the horizon radius and Hawking
temperature show the non-trivial dependence on the dilaton coupling
parameter $a$. This implies that the Hawking temperature increases
exponentially with the coupling parameter $a$.

\vspace{0.25cm}

{\bf Acknowledgment}

\vspace{0.25cm}

We would like to thank the Higher Education Commission, Islamabad,
Pakistan, for its financial support through the {\it Indigenous
Ph.D. 5000 Fellowship Program Batch-IV}.

\end{document}